\documentclass[11pt]{article}
\usepackage{amssymb,latexsym,amsmath}
\title{Gauge-free Electrodynamics}
\author{Parthasarathi Majumdar and Srijit Bhattacharjee\\Saha
Institute of Nuclear Physics\\Kolkata 700 064, India.}
\begin{document}
\maketitle
\begin{abstract}

We propose a reformulation of electrodynamics in
terms of a {\it physical} vector potential entirely free of gauge
ambiguities. Quantizing the theory leads to a propagator
that is gauge invariant by construction in this reformulation, in contrast to
the standard photon propagator. Coupling the theory to a charged Abelian Higgs
field leads at the quantum level to a one loop effective potential which
realizes the Coleman-Weinberg mechanism of
mass generation, thus resolving the issue of 
its gauge dependence. We relate our results to
recent work by Niemi et. al. and Faddeev, where similar strategies are adopted in a
version of the electroweak theory. Other theories
with linear Abelian gauge invariance, like the linearized spin
2 theory of gravity or the antisymmetric tensor field, which may all
be reformulated in terms of physical tensor potentials
without gauge ambiguities, are also discussed briefly. 

\end{abstract}

\section{Introduction}

In vacuum electrodynamics, gauge ambiguities first appear when one solves the
Maxwell-Bianchi identity
\begin{eqnarray}
\partial_{[\mu} F_{\nu \rho]} = 0 \label{bian}
\end{eqnarray} 
in terms of the vector potential $A_{\mu}~:~ F_{\mu \nu}= 2 \partial_{[\mu} A_{\nu]}$;
this solution is unique only upto $U(1)$ gauge transformations: $A_{\mu} \rightarrow
A^{(\omega)}_{\mu} = A_{\mu} + \partial_{\mu} \omega$. This infinite ambiguity
implies that the other Maxwell equation, written out in terms of $A_{\mu}$ 
\begin{eqnarray}
\Box A_{\mu} -\partial_{\mu} \partial \cdot A = J_{\mu} ~\label{maxw}
\end{eqnarray}    
does not yield a unique solution for $A_{\mu}$, for conserved current sources $\partial \cdot J =0$
{\it without a gauge fixing} $G(A)=0$, where the function $G$ may be arbitrary
but must {\it not} be gauge invariant. Clearly, for every choice of $G$, one has $A=A[G]$, and one
can scarcely call any of these $A$s physical. Conventional wisdom relegates the
status of $A$ to that of a subsidiary tool, because of this gauge ambiguity, and
reminds us that only gauge {\it invariant} quantities like $F_{\mu \nu}$ or
Wilson loops $W[C] \equiv \exp i \oint_{C} A \cdot dx$ associated with the
Ehrenberg-Siday-Aharonov-Bohm \cite{esab} phase in quantum mechanics, are truly physical and
measurable.   

In fact, efforts to formulate electrodynamics (and indeed, gauge theories in
general) in terms of gauge {\it invariant} variables have a long history,
starting perhaps with Dirac \cite{dirac55}. A static electron field is characterized
in this approach as a `bare' electron with its own Coulomb field. This is
arrived at using as a phase factor the holonomy of the gauge potential 
$h_C[A](x) = \exp i \int_{C[\infty,x]} A \cdot dx$ for a smooth curve $C$
stretching from asymptopia to the field point $x$. The change in this holonomy operator
under a gauge transformation is exactly compensated by the change in the phase
factor of the bare electron field. Later, Mandelstam \cite{mand} formulated
quantum electrodynamics using such holonomy operators exclusively, without
ever using local gauge potentials themselves. Wilson \cite{wilson} generalized these
holonomy operators, introducing `Wilson loops' as fundamental gauge invariant fields to
formulate Yang-Mills theories, essential to understanding colour
confinement. There have been many efforts in the direction of identifying
gauge invariant variables and formulating gauge theories in terms of
these. See e.g. the recent paper by Ilderton et. al. \cite{ild} which provides
a definitive guide to the literature of the mid-1990s on these efforts, including the
authoritative contribution of Lavelle and McMullan \cite{Mcmullan}. Related to this earlier work,
recently Niemi et. al. \cite{nie} and Faddeev \cite{fadd} have proposed a
gauge invariant description of the Higgs-gauge
sector of standard electroweak theory whereby the Higgs field is given a novel
interpretation as the dilaton in a {\it conformal} curved background. Various
novel aspects of this interpretation have also been reviewed by Ilderton
et. al. \cite{ild}.  

Although similar in spirit to some of these assays in a broad sense, our approach is
distinct in that it is formulated in terms of a {\it local,
physical} vector potential (instead of field strengths) as a {\it
fundamental field variable}. In other words, we
propose an alternative action/field equations as a new starting point rather than attempt to
express the standard gauge theory action in terms of new variables. It is thus a
`gauge-free' approach, rather than one which is based on gauge invariant
functionals of the standard vector potential, including electric/magnetic
fields and Wilson loops.    

To motivate our approach, we begin by noting that, in the standard
formulation, the transverse projection $A^T_{\mu}
= (\delta_{\mu}^{\nu} - \partial_{\mu} \Box^{-1} \partial^{\nu}) A_{\nu}
\equiv {\cal P}^{\nu}_{\mu} A_{\nu}$ is
actually gauge {\it invariant} : $A^{T (\omega)}_{\mu} = A^T_{\mu}$ and also
divergenceless : $\partial \cdot A^T =0$. Observe that this latter property is
{\it independent} of any choice of $\partial \cdot A$ and is by no means
therefore a gauge choice. Further, all physical quantities
like the field strength $F_{\mu \nu}$ and the Ehrenberg-Siday-Aharonov-Bohm phase are
determined by $A^T_{\mu}$ alone and are quite independent of the longitudinal
part $\partial_{\mu} (\Box^{-1} \partial \cdot A) \equiv \partial_{\mu} a_L$
which bears the brunt of the gauge transformation on $A_{\mu} : a_L
\rightarrow a_L^{(\omega)} = a_L + \omega$. Gauge transformations, thus, act
on {\it unphysical} degrees of freedom. Gauge invariance is therefore {\bf not a
physical symmetry}, unlike Poincar\'e invariance in flat spacetime.  

Our approach consists of changing the physical observable that lies at the
heart of electrodynamics: we {\it begin}
with a four-vector potential $A_{{\cal P} \mu}$ as a fundamental {\it local, physical}
field, not as an ambiguous calculational artifact (or even a projection). This field is subject to 
\begin{eqnarray}
\partial \cdot A_{\cal P} =0 ~,\label{trans}
\end{eqnarray} 
as a {\it physical} restriction. This is very similar to the condition $\nabla
\cdot {\vec B} =0$ in standard electrodynamics in three dimensional notation,
imposed on the magnetic field strength ${\vec B}$ as a {\it physical}
condition. In the same sense, eq. (\ref{trans}) is {\it not} a gauge choice, since
there is no gauge ambiguity at all at this level, as the theory is not
based on electromagnetic field strengths (`curvatures') as fundamental
physical variables. The latter are to
be treated as {\it derived} variables. Nor is the photon field $A_{\cal P}$ to
be thought of as a projection of any gauge potential, unlike in some of
the papers in the earlier literature.

Further, we require that $A_{{\cal P}\mu}$ satisfy the field equation 
\begin{eqnarray}
\Box A_{{\cal P}\mu} = - J_{\mu} ~.\label{maxe}
\end{eqnarray}
Eqn.s (\ref{trans}) and (\ref{maxe}) are postulates that {\it define} the fundamental degrees of
freedom and the dynamics of this reformulated gauge-free
electrodynamics. Indeed, they can be derived from the action
\begin{eqnarray}
S[{\bf A}_{\cal P},\Lambda;{\tilde {\bf J}}] &=& \int \left[ -\frac12~\partial_{\mu}
  A_{{\cal P}\nu}~\partial^{\mu}A^{\nu}_{\cal P} + {\tilde J} \cdot A_{\cal P} + \Lambda \partial
  \cdot A_{\cal P} \right] ~ \nonumber \\
&=& \int \left[-\frac12~\partial_{\mu}  A_{{\cal
      P}\nu}~\partial^{\mu}A^{\nu}_{\cal P} + J \cdot A_{\cal P} \right] ~. \label{actn}
\end{eqnarray}
The second line above follows from the first by eliminating the Lagrange
multiplier field $\Lambda$ through its equation of motion and defining
$J_{\mu}$ such that $\partial \cdot J =0$. 

How unique is this `photon' field $A_{\cal P}$, defined as above ? We notice
that indeed, this field does not admit the full $U(1)$ group of local gauge
transformations that the standard electrodynamic gauge potential
does. However, there is a residual ambiguity: the divergencefree constraint
(and also the action) remains invariant under the transformation $A_{{\cal P} \mu}
\rightarrow A_{{\cal P} \mu} + \partial_{\mu} \omega$ for $\omega$ obeying
$\Box \omega = 0$ everywhere. It is obvious though that this ambiguity is of
no consequence for the degrees of freedom coupled to the source. It will
at most affect the {\it homogeneous} solution which is determined solely by
boundary conditions. If we require that $\omega$ is
subject to boundary conditions so that it reduces to a
constant at asymptopia, then the
only globally harmonic function is a constant everywhere. This means that
${\bf A}_{\cal P}$ is effectively uniquely determined by eqn.s (\ref{trans})
and (\ref{maxe}) and hence can well represent the physical photon field. 

As already mentioned, field strengths play no fundamental role in this
construct; they are {\it derived} quantities, given through the standard {\it definition}
$F_{\mu \nu} \equiv 2\partial_{[\mu} A_{{\cal P} \nu]}$, leading to the
Heaviside-Hertz version of the Maxwell equations. Thus, all physical consequences of
standard electrodynamics remain unaltered, while the theoretical framework has
no unphysical degrees of freedom. Further, unlike the standard formulation
where electric-magnetic {\it duality} is deemed to be an important feature,
the gauge-free framework reflects instead electric-magnetic {\it unity}. The
chance discovery of a single elementary magnetic charge is adequate to falsify
this framework in good measure.   

Could this be how Nature actually formulates
electrodynamics, not having the luxury of gauge freedom at her disposal ?  
If so, it must be possible to {\it measure} the physical vector potential without
any prior knowledge of electric and magnetic field strengths. We shall argue
in section 3 that this is indeed possible in this framework. Prior to that in
section 2, the fact that free $A_{{\cal P}\mu}$ fields have two transverse spacelike polarizations will
be seen in section 2 to follow from the defining equations (\ref{trans}),(\ref{maxe}) for
$J_{\mu}=0$. Next, we briefly comment on the Hamiltonian structure of this formulation
of electrodynamics and delineate the difference from the standard
formulation. We then quantize the non-interacting theory and derive the free photon
propagator; it is gauge-inert, as expected. In Section 3, we deal both with charged particles (as already
mentioned) and also charged fields; we couple the theory to charged Higgs fields and
study the Coleman-Weinberg perturbative mass generation mechanism. The
gauge-free nature of the theory eliminates the possibility of gauge dependence
in this mechanism and renders the masses generated to be physical. In the next section (4), we discuss possible
generalization of the gauge-free approach to the linearized spin 2 theory of gravity and
antisymmetric tensor field theory in four dimensional spacetime. We conclude in section 5 with a brief
discussion on Yang Mills fields without Higgs scalars.

\section{Source-free gauge-free electrodynamics}

\subsection{Classical formulation}

Consider now the vacuum sector corresponding to ${\bf J}=0$;
Eq. (\ref{maxe}) reduces to the homogeneous wave equation
\begin{eqnarray}
\Box ~{\bf A}_{\cal P} = 0 \label{waveq}
\end{eqnarray}
which has plane wave like solutions
\begin{eqnarray}
{\bf A}_{\cal P} ~=~ \Re ~{\cal A}~ \exp i \psi ~, \label{solw}
\end{eqnarray}
where, we assume that the amplitude ${\cal A}$ is a slowly-varying
one form field on spacetime, while the phase function $\psi$ varies much more
rapidly; in other words, schematically
\begin{eqnarray}
|\partial_{\mu}\psi| >> |\partial_{\mu} {\cal A}| ~\label{eik}
\end{eqnarray}
for all components. Substituting the ans\"atz (\ref{solw}) into
(\ref{trans}) and (\ref{waveq}), and defining ${\bf k} \equiv
-i d~\psi$ one obtains in the approximation (with slight abuse of notation)
\begin{eqnarray}
{\bf k} \cdot {\bf A}_{\cal P} ~=~ 0 ~=~ k^2 \label{oms}
\end{eqnarray}
Thus ${\bf A}_{\cal P}$ as a vector field  is orthogonal in spacetime to the
null vector field ${\bf k}$. As we argue below, this has the important
consequence that ${\bf A}_{\cal P}$ must be spacelike everywhere, or null
in regions of spacetime where it is parallel to ${\bf k}$.

Consider the null tetrad basis of flat spacetime, given by the four null
vectors ${\bf l}~,~{\bf n}$ which are real, and the complex conjugate pair
${\bf m}~,~{\bar {\bf m}}$, obeying the conditions
\begin{eqnarray}
{\bf l} \cdot {\bf n} ~=~ 1~,~{\bf m} \cdot {\bar{\bf m}}~=~-~1
\end{eqnarray}
with all other scalar products vanishing. Without any loss of generality, the
null tetrad basis may be so chosen that the basis vector ${\bf n}$ is parallel
with the wave vector ${\bf k}$ defined above. Expanding the transverse
potential ${\bf A}_{\cal P}$ in this basis
\begin{eqnarray}
{\bf A}_{\cal P}~=~A_{\cal P}^{(n)}~{\bf l} ~+~A_{\cal P}^{(l)}~{\bf n}~-~
A_{\cal P}^{({\bar m})}~{\bf m}~-~A_{\cal P}^{(m)}~{\bar {\bf m}} ~, \label{expn}
\end{eqnarray}
the first of the equations (\ref{oms}) immediately implies that the component
$A_{\cal P}^{(n)} = {\bf n} \cdot A_{\cal P} = 0$. Consequently, the
squared norm of ${\bf A}_{\cal P}$ (considered as a vector field in spacetime) is given by
\begin{eqnarray}
{\bf A}_{\cal P}^2 ~=~ - ~2 |A_{\cal P}^{(m)}|^2 ~<~ 0, \label{ass}
\end{eqnarray}
implying that this potential is everywhere spacelike. Furthermore, {\it the integral
curves of the potential 4 vector field are closed curves in
spacetime.} It is quite remarkable that even though this was not put in, there
are no closed {\it timelike} integral curves of the vector potential.

Clearly, the component $A_{\cal P}^{(l)}$, even though not
necessarily zero in an
arbitrary Lorentz frame, may indeed be made to vanish by a suitable choice of
a frame. In this special frame, the only non-vanishing components are
the complex combination $A_{\cal P}^{(m)}$. It is now natural to identify
these components with the two spatial degrees of transverse polarization that
electromagnetic waves in vacuum are known to have. These correspond in the
quantum description to the two
helicity states of helicity $\pm 1$ of a massless vector field in Minkowski
spacetime. Note that the existence of the two spatial polarizations does
not involve any gauge choice; it does however involve a choice of a Lorentz
frame. One further note of interest is that the complex polarization $A_{\cal
P}^{(m)}$ basically spans a 2-sphere (for free photon fields of a given intensity) which may be identified
with the Poincar\'e sphere.   

\subsection{Hamiltonian structure}

It is straightforward to compute the Hamiltonian corresponding to the action
(\ref{actn}); one begins with the canonical momenta given by 
\begin{eqnarray}
\Pi_0 = \Lambda - \partial_0 A_{{\cal P} 0}~,~ \Pi_{\alpha} = - \partial_0
A_{\cal P \alpha} , \label{mom}
\end{eqnarray}
and computes the Hamiltonian  
\begin{eqnarray}
H = \int d^3x ~\left[ \Pi_0 ~\partial_0 A_{\cal P}^0 + \Pi_{\alpha} ~\partial_0
  A_{\cal P}^{\alpha} \right] - L ~, \label{ha}
\end{eqnarray} 
where $L$ is the free field Lagrangian corresponding to the action
(\ref{actn}). Using the constraint (\ref{trans}) written out in terms of the
canonical momenta, and eliminating the auxiliary field $\Lambda$, one gets
\begin{eqnarray}
H = \frac12 ~\int d^3x ~\left[ {\vec \Pi}^2 + (\nabla {\vec A}_{\cal
        P})^2 + (\nabla \cdot {\vec A}_{\cal P})^2 - (\partial_{\alpha}
  A_{\cal P}^0)^2 - 2 \Pi_0 ~\nabla \cdot {\vec A}_{\cal P} \right]
~. \label{hamil}
\end{eqnarray}

The differences from the standard formulation are as follows: 
\begin{itemize}
\item The component $A_{\cal P}^0$ is in our case no longer a Lagrange
multiplier field with vanishing canonical momentum. However, its canonical
momentum is determined in terms of other degrees of freedom because of the
relation (\ref{trans}).
\item The Hamiltonian (\ref{hamil}) is clearly bounded from below, keeping in
mind the spacelike nature of the photon polarizations on shell. It is however
{\it not} identical to the form found in the standard formalism where it is
proportional to the sum of squares of the electric and magnetic fields.
\item There are no constraints, either first or second class.
\end{itemize}

\subsection{Functional quantization}

The quantum theory of a free transverse vector field can arrived at using
the functional integral formulation. In contrast to the
standard approach where the Faddeev-Popov prescription is usually
employed to integrate out the gauge degrees of freedom, no gauge
fixing need be employed here. However, since the functional integral
describing the vacuum-to-vacuum amplitude is over all configurations
of the vector field ${\bf A}_{\cal P}$, the transversality constraint
(\ref{trans}) must be directly inserted into the integral to ensure
that the integral is only over transverse field configurations. Recall
that this is {\it not} a gauge condition; it is instead a constraint
defining the physical photon degrees of freedom.

The relevant vacuum-to-vacuum amplitude (in presence of a transverse
source) is given by
\begin{eqnarray}
Z[{\bf J}] ~&=&~ \int {\cal D}{\bf A}_{\cal P}~ \exp i \left(~ \frac12
~A_{\cal P}^{\mu}~ \Box~A_{{\cal P} \mu}   ~+~ \int d^4x~{\bf J}
\cdot {\bf A}_{\cal P} \right) ~ \delta[\partial_{\mu}
~{\bf A}_{\cal P}^{\mu}] \nonumber \\
& ~=~& \int {\cal D}{\cal S}~{\cal D}{\bf A}_{\cal P}~\exp i~\int d^4x~ \left [ \frac12
~A_{\cal P}^{\mu}~ \Box~A_{{\cal P} \mu} ~+~ ( J_{\mu} ~-~ \partial_{\mu} ~{\cal
  S}) ~A_{\cal P}^{\mu} \right] ~. ~\label{vacam}
\end{eqnarray}
In the second line of (\ref{vacam}) we have introduced an auxiliary
scalar field ${\cal S}$ which acts as the Lagrange multiplier for the
physical constraint (\ref{trans}). The integral over ${\bf A}_{\cal P}$  is
easily done, yielding
\begin{eqnarray}
Z[{\bf J}] &~=~&(\det ~\eta_{\mu \nu}\Box)^{-1/2}~ \int {\cal D}{\cal S} ~ \exp \frac{-i}{2} \int
d^4x~d^4y~[\partial_{\mu}^x {\cal S}(x)~-~ J_{\mu}(x)]~ \nonumber \\
~& \cdot &~{\cal G}(x-y)~\eta^{\mu \nu}~[\partial_{\nu}^y {\cal S}(y) ~-~J_{\nu}(y)]~, \label{integr}
\end{eqnarray}
where, ${\cal G}(x-y)$ is the d'Alembertian Green's function. A series of
partial integrations and using the transversality of the current density ${\bf
J}$, and also identities like $\partial_x {\cal G}(x-y) = - \partial_y {\cal
D}(x-y)$ leads to the simple expression
\begin{eqnarray}
Z[{\bf J}] &~=~& (\det ~\eta_{\mu \nu} \Box)^{-1/2} ~\exp -\frac12 ~i\int d^4 x d^4 y
J^{\mu}(x) \Delta_{\mu \nu}(x-y) J^{\nu}(y) ~ \nonumber \\
~&\cdot &~ \int {\cal D}{\cal S} ~\exp -\frac{i}{2} \int d^4x \,{\cal S}^2(x) ~, \label{fctalint}
\end{eqnarray}
where, $\Delta_{\mu \nu}(x-y) \equiv \eta_{\mu \nu}~{\cal G}(x-y)$. The
integral over ${\cal S}$ is a trivial Gaussian producing a constant (albeit
infinite) independent of ${\bf J}$. It will thus cancel out, along with the
prefactor $det^{-1/2} \Box$ in eq. (\ref{integr}) from the generating
functional for connected Green's functions $W[{\bf J}] = -i \log Z[{\bf J}]$,
and all vacuum expectation values.

It is now straightforward to extract the free photon propagator from
eq. (\ref{fctalint}):
\begin{eqnarray}
{\cal G}_{\mu \nu}(x-y) &~\equiv~& \frac12 ~{\delta^2 W[{\bf J}] \over \delta
 J^{\mu}(x) ~\delta J^{\nu}(y) }\Bigg|_{{\bf J}=0} \\
&~=~& \eta_{\mu \nu} {\cal G}(x-y) ~=~\int {d^4p \over (2\pi)^4}~{i~\eta_{\mu
  \nu} \over p^2 ~+~i \epsilon} ~. \label{photpr}
\end{eqnarray}
Clearly, this propagator does not possess any gauge
artifacts.

We close this section with an observation: even though the
physical photon field is
spacetime transverse, its propagator is {\it not} the same as the {\it
Landau} gauge propagator of standard QED. It is what in the
standard formulation would correspond to the {\it Feynman} gauge. This is different
from the Landau-Lorentz gauge condition which formally resembles our
divergence free constraint (\ref{trans}). In the
standard approach, one adds to the free Maxwell action the gauge
fixing term $(1/2\alpha) (\partial \cdot {\bf A})^2$ corresponding to
the so-called $\alpha-$gauges. The Feynman gauge corresponds to the
choice $\alpha=-1$ and the Lorentz-Landau gauge to the limit $\alpha
\rightarrow 0$. In our gauge-free formulation, spacetime
divergencelessness is {\it not} a matter of choice, it is a defining
feature of what we mean by electromagnetism. Finally, note also that the free
photon propagator falls off as $1/p^2$ for large momentum, as is expected for
a {\it local} field.  

\section{Gauge-free electrodynamics with sources}

\subsection{Charged particle dynamics in external ${\bf A}_{\cal P}$ fields}

The action in this case is given (for a single point charge) by
\begin{eqnarray}
S[{\bf x}, {\bf A}_{\cal P}] = \int d\tau \left[ -m (u^2)^{1/2} + e
  {\bf u} \cdot {\bf A}_{\cal P} \right] ~, \label{char}
\end{eqnarray}
where, ${\bf u} \equiv d {\bf x} / d\tau$ is the 4-velocity of the charge
$e$. Rather than the Euler-Lagrange equation which leads immediately to the
relativistic Lorentz force equation, consider the Hamilton-Jacobi equation
appropriate to this action \cite{llctf}. The 4-momentum
${\bf P}$ is defined in terms of the kinematical momentum  ${\bf p} \equiv m{\bf u}$ as $ {\bf P} \equiv
{\bf p} +e {\bf A}_{\cal P}$. To write down the Hamilton-Jacobi equation, ${\bf
p}$ is replaced in the mass-shell condition $P^2 \equiv ({\bf p} + e{\bf
A}_{\cal P})^2 = m^2$ : $p_{\mu} \rightarrow \partial_{\mu} S$ so that one
obtains
\begin{eqnarray}
\left ( \partial_{\mu} S + e~A_{{\cal P} \mu} \right)\left ( \partial^{\mu} S
  + e~A_{\cal P}^{ \mu} \right) = m^2 ~. \label{haja}
\end{eqnarray}
Given an external ${\bf A}_{\cal P}$ configuration, the object is to solve
(\ref{haja}) first for $S$ as a functional of ${\bf A}_{\cal P}$. The
action-angle variable formalism is then used to determine the particle trajectory as
a functional of ${\bf A}_{\cal P}$. Nowhere does one need to invoke field
strengths to determine the motion of the charge. For known particle
trajectories, it is a mathematical exercise to determine the external ${\bf
  A}_{\cal P}$ which produces those trajectories. This vector potential is
thus in-principle measurable as a physical quantity without any gauge ambiguity.      
 
\subsection{Charged matter fields}

All charged matter fields are complex fields ${\Phi}$
such that they can be `radially' decomposed : ${\Phi} = {\phi} \exp i\theta$ where 
${\phi}$ carries all the spin degrees of freedom of ${\Phi}$ and the
phase field $\theta$ is a scalar field which appears in the action only
through its first order derivative $\partial \theta : S[\Phi] =
S[\phi, \partial\theta]$. The gauge-free prescription for coupling the
gauge-free vector potential ${\bf A}_{\cal P}$ to $\Phi$ is exceedingly simple
: leaving $\phi$ as it is in the action, simply replace $\partial \theta
\rightarrow \partial \theta - e A_{\cal P}$, so that $S[\Phi] \rightarrow
S[\phi, \partial \theta - e A_{\cal P}] + S_{free} [A_{\cal P}]$. Recall of
course that the gauge-free ${\bf A}_{\cal P}$ is subject to the
4-divergencelessness constraint (\ref{trans}). The interaction with matter for
this vector potential is merely to add a {\it physical longitudinal} part to
it so that potentially it can now turn massive even in the weak coupling
limit, depending upon the form of $S[\Phi]$. An example of this is the Abelian
Higgs model of scalar electrodynamics.  

\subsection{Abelian Higgs Model}

A charged scalar admits the radial decomposition $\phi = (\rho/\sqrt{2}) \exp i\Theta$
where $\rho$ and $\Theta$ are both to be treated as physical fields. With this
decomposition, the action of the complex scalar field appears as (suppressing
obvious indices) 
\begin{eqnarray}
S_0[\rho, \Theta] = \int d^4 x \left[\frac12 (\partial \rho)^2 + \frac12 \rho^2 (\partial
  \Theta)^2 - V(\rho) \right] ~. \label{comsc}
\end{eqnarray}
This action (\ref{comsc}) is invariant under the global $U(1)$ transformations $ \rho
\rightarrow \rho~,~ \Theta \rightarrow \Theta + \omega$ where $\omega$ is a
real constant. 

Following our prescription above, coupling to the physical electromagnetic
vector potential is obtained through the action (dropping obvious indices)
\begin{eqnarray}
S[\rho,\Theta,{\bf A}_{\cal P}] = \int d^4x \left[ \frac12 (\partial
\rho)^2 + \frac12 e^2 \rho^2 ({\bf A}_{\cal P} - e^{-1}\partial \Theta)^2
~- \frac12 (\partial A_{\cal P})^2 - V(\rho) \right], \label{abh2}
\end{eqnarray}
where $V(\rho)$ is the scalar potential, and ${\bf A}_{\cal P}$
obeys the divergenceless constraint (\ref{trans}). 
It is interesting that the phase field $\Theta$
occurs in the action only through the combination ${\bf A}_{\cal P} - e^{-1}
d\Theta$; this implies that the shift $\Theta \rightarrow \Theta +
const.$ is still a symmetry of the action. However, since there is no
canonical kinetic energy term for
$\Theta$, it is hard to associate a propagating degree of freedom with
$\Theta$. Indeed, if one first makes a field redefinition
\begin{eqnarray}
Y_{\mu} ~\equiv~ A_{{\cal P} \mu}~ -~e^{-1}~ \partial_{\mu}~\Theta ~. \label{ugau}
\end{eqnarray}
the $\Theta$ can be completely absorbed into the new vector field ${\bf
  Y}_{\mu}$, appearing only in the constraint which replaces (\ref{trans}) 
\begin{eqnarray}
\partial \cdot Y = - \Box \Theta ~. \label{try}
\end{eqnarray}
This implies that ${\bf Y}$ has three {\it physical} polarizations rather than the two that
${\bf A}_{\cal P}$ had. However, this does not immediately imply that ${\bf
Y}$ has acquired a mass. Upon eliminating $\Theta$ through the constraint
(\ref{try}), eq. (\ref{abhm}) assumes the form
\begin{eqnarray}
S[\rho,{\bf Y}] = \int d^4x \left[ \frac12 (\partial
\rho)^2 + \frac12 Y^a \left( (\Box + e^2\rho^2) \eta_{ab}
  -\partial_a \partial_b \right) Y^b - V(\rho) \right], \label{abhm}
\end{eqnarray}
This is the gauge-free Abelian Higgs model. 

One can now think of two kinds of scalar potentials $V(\rho)$: one for
which the minimum of the potential $\langle \rho \rangle =0$ and the
other for which the minimum lies away from the origin $\langle \rho
\rangle = \rho_0 \neq 0$. It is this second case which is of interest to us.
If $V(\rho)$ has a minimum at $\rho = \rho_0 \neq 0$
one now also defines $\rho \rightarrow \rho +
\rho_0$, it is easy to see that the ${\bf Y}$ acquires a mass
$m_{\bf Y}^2 = e^2 \rho_0^2$ while the $\rho$ also acquires a mass
$m_{\rho}^2 = V''(\rho_0)$. This is precisely the manner in which a {\it
physical} longitudinal degree of freedom conjoins the photon field to produce
a massive vector boson. In doing so, the new vector
potential ${\bf Y}$ is no longer subject to the transversality constraint (\ref{trans}). It
thus has one degree of freedom more than the ${\bf A}_{\cal P}$. Observe that
the Higgs phenomenon of mass generation {\bf did not involve any symmetry
breaking at all}, reminding us of Elitzur's theorem \cite{elit} proved for QED on a
cubic lattice. The vacuum expectation value $\rho_0 \equiv \langle \rho \rangle$ does not break any
continuous symmetry at all. {\it The Higgs mechanism is a gauge-free mechanism
of mass generation, involving neither symmetry breaking of any sort, nor
unphysical particles in the spectrum}. 

Before closing this subsection, we point out that this aspect of the phase
field attaching itself to the photon field as a {\it physical}
longitudinal piece, is not confined to charged scalar fields. Consider for
instance a free charged Dirac field given by the action 
\begin{eqnarray}
S[\psi] = \int d^4 x~ {\bar \psi} (i \gamma \cdot \partial -m) \psi
~.\label{frd1}
\end{eqnarray}
Performing the `radial decomposition' $\psi = \chi \exp i\theta$ this reduces
to 
\begin{eqnarray}
S[\chi, \theta] = \int d^4 x \left( ~{\bar \chi}(i \gamma \cdot \partial -m)
  \chi - {\bar \chi} \gamma \cdot \partial \theta \chi~ \right) ~. \label{frd2}
\end{eqnarray} 
This action is of course invariant under the global $U(1)$ transformations
$\chi \rightarrow \chi~,~ \theta \rightarrow \theta + \omega$ for a constant
$\omega$. Employing our prescription above for coupling this field to the physical
electromagnetic vector potential, we notice that the action now reads
\begin{eqnarray}
S[\chi, \theta] = \int d^4 x \left( ~{\bar \chi}(i \gamma \cdot \partial -m)
  \chi - {\bar \chi} \gamma 
\chi \cdot (\partial \theta - e A_{\cal P})~ \right) ~. \label{frd3}
\end{eqnarray}
It is obvious from the above that under any interaction, the vector potential
is {\it poised} to pick up a physical longitudinal piece ($\partial \theta$)
corresponding to the `charge mode'. However, in this case there is no
mechanism (at tree level) of mass generation due to the absence of a `seagull'
term. But this could be an artifact of weak coupling. In the 1+1 dimensional
quantum electrodynamics model analyzed half a century ago by Schwinger \cite{schwi}, the
photon field does pick up a manifestly gauge invariant mass as an exact dynamical result.

\subsection{Gauge-free scalar QED: Coleman-Weinberg Mechanism}

The Coleman-Weinberg mechanism \cite{cw} is a radiative mechanism whereby a
scalar electrodynamics theory with massless photons and charged scalar bosons, changes
its spectrum due to perturbative quantum corrections. Both the neutral
component of the scalar boson and the vector boson acquire physical masses given by the
parameters of the theory. In its incipient formulation, the mechanism has been
shown to be gauge-dependent \cite{jacki}, thereby casting doubt on its
physicality. Using the gauge free reformulation given above, we compute in
this section the one loop effective potential of the theory, and argue that
the effect is physical at this level.

The action for the theory is already given above (eq. (\ref{abhm})), with the
choice $V(\rho) = (\lambda/4!) \rho^4$. Following
\cite{cw}, the theory is
quantized using the functional integral formalism. In the standard formulation
of QED, one needs to resort to the Faddeev-Popov technique of gauge fixing and
extracting the infinite volume factor associated with the group of gauge
transformations, from the vacuum persistence amplitude (generating functional
for all Green's functions), in
order that this amplitude does not diverge upon integrating over gauge
equivalent copies of the gauge potential. In the gauge free approach here, this
technique is not necessary. The integration over the transverse gauge
potential is, of course, restricted to configurations that obey the spacetime
transversality condition (\ref{trans}). Since the integration variables are
unambiguous, the task, at least at the one loop level, is simpler.

The generating functional is thus given by
\begin{eqnarray}
Z[J, J', {\bf J}] &=& \int {\cal D}\rho ~{\cal D}\Theta ~{\cal D}{\bf A}_{\cal
  P}  ~\exp {i \over \hbar} \left[~S[ \rho,\Theta,{\bf A}_{\cal P} ]
+ \int d^4x(J \rho + J' \Theta + {\bf J} \cdot {\bf A}_{\cal P})~ \right]
\nonumber \\
&&~ \cdot \delta[~\partial_{\mu}~A^{\mu}_{\cal P}~] ~. \label{vpam}
\end{eqnarray}
Here, the integration measures ${\cal D} \rho = \Pi_x d\rho(x)~,~{\cal D}{\bf
A}_{\cal P} = \Pi d{\bf A}_{\cal P}$, but the remaining measure ${\cal
D}\Theta = Det \rho^2 \Pi_x d\Theta(x)$. The extra factor of $Det \rho^2$
can be seen to arise if one begins with the generating functional first
expressed as functional integrals over a complex scalar field and its complex
conjugate. Alternatively, one can obtain the configuration space functional
integral starting with the functional integral over phase space. Integration
over the momentum conjugate to $\Theta$ produces the same factor \cite{senj}.

Indeed, it is a similar factor which has been interpreted in \cite{fadd} as
representative of a background spacetime which is {\it conformally} flat,
rather than flat, with the `radial' component of the Higgs field $\rho$ playing the
role of the conformal mode. In \cite{nie}, a slightly different interpretation is
given of this radial Higgs field as a {\it dilaton} field. Formally, there is
indeed novelty in both interpretations. However, when perturbative effects are
included, at least at the one loop level, such interpretations will be seen to
be in need of modification to account for scaling violations due to renormalization.  

The effective action $\Gamma[\Phi]$ which is the generating functional for one particle
irreducible diagrams (1PI), is generically defined as usual through the Legendre
transformation
\begin{eqnarray}
\Gamma[\Phi] ~&=&~ W[{\cal J}] ~-~\int d^4 x~{\cal J} \cdot \Phi~ \nonumber \\
\Phi ~&=&~ {\delta W[{\cal J}] \over \delta {\cal J}} ~,
\end{eqnarray}
where, we have collectively labeled all fields as $\Phi$ and the sources as
${\cal J}$, and $W[{\cal J}]$, we recall, is the
generating functional of connected Green's functions. The task is to
compute $\Gamma[\Phi]$ to ${\cal O}(\hbar)$ with a view to eventually obtaining
the one loop effective potential defined by the relation
\begin{eqnarray}
V_{eff}(\phi_0) ~\equiv ~-~ \Gamma(\Phi)|_{\Phi=\phi_0} ~\left(\int d^4x \right)^{-1} ~ ,
\label{veff}
\end{eqnarray}
where, $\phi_0$ are spacetime independent. Observe that $V_{eff}(\phi_0)$ is
the generating functional for 1PI graphs with vanishing external
momenta. Even though the scalar potential is classically scale invariant,
a mass scale is generated through renormalization in the quantum theory,
which breaks this scale invariance. The effective potential may thus have
a minimum away from the origin in $\rho$-space, defined in terms of the
renormalization mass scale, which, in turn, relates to values of the dimensionless
physical parameters of the theory (dimensional {\it transmutation} \cite{cw}).

Instead of evaluation of the functional integral over the $\Theta$ and ${\bf A}_{\cal
P}$ fields, we make a change of basis to $\Theta$ and ${\bf Y}$ via
(\ref{ugau}) and make use of
the action (\ref{abh2}) which is independent of $\Theta$. The latter appears
only in the constraint which now becomes a statement of non-transversality in
spacetime of the ${\bf Y}$ field. $\Theta$ can be simply integrated out,
leaving behind a field-independent normalization which we set to unity.
The integration over $\rho$
involves a saddle-point approximation around a field $\rho_c$ which may be
called a `quantum' field, since it is the solution of the classical
$\rho$-equation of motion augmented by ${\cal O}(\hbar)$ corrections. With
no gauge ambiguities anywhere, there is no question of gauge fixing; functional
integration over the physical vector potential ${\bf Y}$ can be performed
straightforwardly.

Following ref. \cite{jacki}, the one loop effective action is given
schematically by
\begin{eqnarray}
\Gamma^{(1)}~[\rho_c] ~=~ S[\rho_c,0,0] ~-~ i\hbar~Z^{(1)}[\rho_c] ~,
\label{olga}
\end{eqnarray}
where,
\begin{eqnarray}
Z^{(1)}[\rho_c]\equiv  \int {\cal D} \rho {\cal D}{\bf Y} ~\exp {i \over 2\hbar}
\left[~\int d^4xd^4y~\rho(x) {\cal M}_{\rho\rho}(x,y)~\rho(y) + Y^{\mu}(x) {\cal
M}_{Y_{\mu}Y_{\nu}}(x,y) Y^{\nu}(y)~ \right]~, \label{zone}
\end{eqnarray}
with, generically,
\begin{eqnarray}
{\cal M}_{AB}(x,y) ~ \equiv ~ \left({\delta^2 S[\Phi] \over \delta
\Phi_A(x)~\delta \Phi_B(y)}\right)_{\Phi=\rho_c,0,0}~. \label{emm}
\end{eqnarray}
Since our object of interest is the one loop effective potential, we restrict
ourselves to a saddle point $\rho_c$ which is spacetime independent.
The matrices ${\cal M}$ turn out to be diagonal in field space for the purpose of a one loop
computation, with entries
\begin{eqnarray}
{\cal M}_{\rho \rho}~ &=&~ - \left(~\Box ~+~ {\lambda \over 2} \rho_c^2
  ~\right) ~\delta^{(4)}(x-y) \nonumber \\
{\cal M}_{Y_{\mu} Y_{\nu}}~ &=&~\left[ \eta_{\mu \nu}~\left(
  ~\Box~+~e^2~\rho_c^2\right) - \partial_{\mu} \partial_{\nu} \right]\delta^{(4)}(x-y) ~.
\end{eqnarray}
One obtains easily
\begin{eqnarray}
Z^{(1)}[\rho_c]~&=&~ \left(~Det \left[ ~{\cal M}_{\rho \rho} {\cal M}_{YY}
    ~\right] \right)^{-1/2}~ ,\label{zloop}
\end{eqnarray}

The functional determinants are evaluated in momentum space following
\cite{jacki}, and one obtains for the one loop
effective potential, using eq. (\ref{veff}), the expression
\begin{eqnarray}
V_{eff}(\rho_c) &=& \frac{1}{4!} \lambda \rho_c^4~+~\hbar
\int d^4k~\log \left [ \left(-k^2 + e^2 \rho_c^2 \right)^{3/2} \left
(-k^2 + \lambda \rho_c^2 \right)^{1/2} \right] \nonumber \\
&+& \frac12 B \rho_c^2 ~+~ \frac{1}{4!} C \rho_c^4 ~,
\end{eqnarray}
where $B$ and $C$ are respectively the mass and coupling constant
counterterms. The momentum integral is performed with a
Lorentz-invariant cut-off $k^2 = \Lambda^2$, yielding
\begin{eqnarray}
V_{eff}(\rho_c) &=& \frac{1}{4!}\rho_c^4~+~\frac12 B \rho_c^2 ~+~ \frac{1}{4!}
C \rho_c^4 \nonumber \\
&~+~& {\hbar \rho_c^2 \Lambda^2 \over 32 \pi^2} (\frac12 \lambda + 3
e^2) \nonumber \\
&+& {\hbar \rho_c^4 \over 64\pi^2} \left[\frac14 \lambda^2
  \left(\log {\lambda \rho_c^2 \over 2 \Lambda^2} - \frac12 \right)
+ 3 e^4 \left(\log {e^2 \rho_c^2 \over \Lambda^2} - \frac12 \right)
\right]
\end{eqnarray}

We remark here that in these manipulations, a $\exp (-\log \rho^2)$ term is
generated in the one loop partition function, which cancels {\it exactly}
against an identical term $Det \rho^2$ arising in the formal measure as
discussed after eq. (\ref{vpam}). This is precisely the point that was made
earlier: the interpretation of that extra local factor in the formal
functional measure as some sort of conformal mode in a conformally flat
background is subject to some modification at the one loop level, since that factor is
{\it eliminated} by a one loop contribution to the partition function. This
has been anticipated in ref. \cite{rys} where an attempt has been made to give
an alternate interpretation in terms of a `gauge-dependent gravity'. Perhaps one
can use compensator fields to account for this loss of scale invariance due
to renormalization effects, in order to resurrect the novel interpretation
proposed in \cite{nie}, \cite{fadd}. 
  
The mass and coupling constant renormalizations $B$ and $C$ are fixed through
the renormalization conditions
\begin{eqnarray}
{d^2 V \over d\rho_c^2}\Bigg|_{\rho_c=M} &=& 0 \\
{d^4 V \over d\rho_c^4}\Bigg|_{\rho_c=M} &=& \lambda
\label{renc}
\end{eqnarray}
leading to the renormalized one loop effective potential
\begin{eqnarray}
V_{eff}(\rho_c) &=& {\lambda \over 4!} \rho_c^4 + \rho_c^2 M^2 \left[-{\lambda
    \over 4} + {9 \over 32\pi^2} (3e^4 + \frac12 \lambda^2) \right] \\
&+& \left({3e^4 \over 64 \pi^2} + {\lambda^2 \over 256 \pi^2} \right) \rho_c^4
\left[\log {\rho_c^2 \over M^2} - {25\over 6} \right]~.
\label{rveff}
\end{eqnarray}

The potential has an extremum at $\rho_c = \langle \rho \rangle(M)$ leading
eventually to the ratio of the squared masses of the Higgs boson to the photon
\begin{eqnarray}
{m_H^2 \over m_A^2} ~=~ \frac{1}{e^2} ~\left[\frac13 \lambda -\left({3e^4
      \over 8 \pi^2} + {\lambda^2 \over 32\pi^2} \right) \log {\langle \rho
      \rangle^2 \over M^2 } -{e^4 \over \pi^2} - {\lambda^2 \over 12\pi^2} \right]
\label{masr}
\end{eqnarray}
The derivation of the mass ratio of the Higgs mass to the photon seemingly
went through without any gauge fixing, since all fields being functionally
integrated over are {\it physical} fields without any gauge ambiguity. The
result (\ref{masr}) is thus a `physical' result in this toy model where the
photon acquires a mass. Notice that unlike in the original Coleman-Weinberg
paper, we did not make an approximation of choosing $\lambda \sim e^4$, to
drop terms of $O(\lambda^2)$. Thus, even though our result agrees with the
earlier papers qualitatively, there are significant quantitative
differences. However, the point in this section is not so much the result of
the computation of the mass ratio, but the observation that the effect
is physical and not a gauge artifact.

\section{Generalizations}

The projection operator ${\cal P}$ defined earlier can be used to project out the physical
(i.e., gauge invariant) part of the spin 2 linearized graviton field and
the Kalb-Ramond antisymmetric second rank gauge potential \cite{ kalb}, as we now proceed to
show. This leads us immediately to formulating the theory of these tensor
fields in terms of gauge-free graviton and antisymmetric tensor fields. 

\subsection{Graviton field}

The graviton field is defined in terms of spin two fluctuations about
Minkowski spacetime
\begin{eqnarray}
g_{\mu \nu} ~=~\eta_{\mu \nu} ~+~h_{\mu \nu} ~. \label{sp2}
\end{eqnarray}
If the Einstein-Hilbert action is expanded in powers of the spin 2 fluctuations
$h_{\mu \nu}$ upto bilinear terms, the effective action is invariant under
linearized infinitesimal coordinate (gauge) transformations 
\begin{eqnarray}
h_{\mu \nu} ~\rightarrow~ h_{\mu \nu} + 2 \partial_{(\mu} \xi_{\nu)}
~. \label{linco}
\end{eqnarray}

Consider now the double projection on these spin 2 fluctuations
\begin{eqnarray}
h^T_{\mu \nu} ~\equiv {\cal P}_{\mu}^{\lambda}~{\cal
  P}_{\nu}^{\rho}~h_{\lambda \rho}
\end{eqnarray}
with ${\cal P}$ defined as earlier. It is easy to verify that under the
linearized coordinate transformation (\ref{linco}), $h^T_{\mu \nu}$ is {\it
invariant}. Further, it satisfies the spacetime transversality condition
\begin{eqnarray}
\partial^{\mu}~h^T_{\mu \nu}~=~ 0 ~. \label{transh}
\end{eqnarray}

The linearized equation of motion for the graviton field is given by
\begin{eqnarray}
{\cal G}_{\mu \nu} &\equiv& \frac12 \left( \partial_{\mu} \partial_{\nu} h ~+~ \Box
h_{\mu \nu} \right) - \partial_{\rho} \partial_{(\mu} h^{\rho}_{\nu)}
\nonumber \\
&-& \frac12 \eta_{\mu\nu}\left( \Box h - \partial_{\rho} \partial_{\lambda} h^{\rho \lambda}
\right) \nonumber \\
&=& 8 \pi G T_{\mu \nu} ~, \label{lineq}
\end{eqnarray}
where, $h \equiv h_{\mu} ^{\mu}$. In terms of the projected tensor field
$h^T_{\mu \nu}$, this equation reduces to
\begin{eqnarray}
{\cal G}_{\mu\nu} ~\equiv ~ \frac12 \Box \left ( h^T_{\mu \nu} ~-~ {\cal P}_{\mu
\nu} h^T \right)   ~.
\end{eqnarray}
Defining
\begin{eqnarray}
{\bar h}^T_{\mu \nu}  \equiv h^T_{\mu \nu} - {\cal P}_{\mu \nu} h^T
~, \label{htbar}
\end{eqnarray}
the linearized equation reduces to the inhomogeneous d'Alembert wave equation
\begin{eqnarray}
{\cal G}_{\mu \nu} ~=~ \frac12 \Box {\bar h}^T_{\mu \nu} = 8\pi G T_{\mu \nu}
~. \label{daal}
\end{eqnarray}
The field ${\bar h}^T_{\mu \nu}$ is also spacetime transverse and manifestly
gauge invariant just like $h^T_{\mu \nu}$. However, note that it is {\it not}
traceless : ${\bar h}^T \neq 0$. 

With this as motivation, it is now possible to define the physical graviton
field $h_{\cal P \mu \nu}$ which obeys 
\begin{eqnarray}
\partial_{\mu} ~ h^{\mu \nu}_{\cal P} &=& 0 \nonumber \\
\Box h_{\cal P \mu \nu} &=& 8 \pi G T_{\mu \nu}  ~. \label{gra}
\end{eqnarray}
How unique is this physical graviton field ? If we make the standard linear
coordinate gauge transformation for graviton fields discussed above, i.e., $\delta
h_{\cal P \mu \nu} = 2 \partial_{(\mu} \xi_{\nu)}$, we find, from
eqn. (\ref{gra}) that the gauge
function $\xi^{\mu}$ must satisfy the equation $\Box \xi_{\mu}
+ \partial_{\mu}\partial \cdot \xi =0$ which does not appear to have any
nontrivial solution ! Our physical graviton field is thus unique. One can
show, following the
procedure adopted for the photon field, that there is a Lorentz frame in which
the free graviton field $h_{\cal P \mu \nu}$ has only {\it two} spatial
polarizations. This particular frame defines the so-called {\it
transverse-traceless} polarization degrees of freedom relevant for
gravitational wave detection. 

\subsection{Kalb-Ramond two form potential}

The Kalb-Ramond two form potential ${\bf B}$ has a field strength ${\bf H} =
d{\bf B}$ which is clearly invariant under the gauge transformation ${\bf B}
\rightarrow {\bf B} + d\Lambda$ for any one form field $\Lambda$. Construct
now the projected two form field ${\bf B}^T \equiv {\cal P} \otimes {\cal P}
{\bf B}$. Since $ {\cal P} d f=0~ \forall f$, under the
gauge transformation of ${\bf B}$, ${\bf B}^T \rightarrow {\bf B}^T + {\cal
P}\otimes {\cal P} d \Lambda = {\bf B}^T$. Further, in a coordinate system,
\begin{eqnarray}
\partial _{\mu} ~B^{T \mu \nu}~=~ 0 \label{trkr}
\end{eqnarray}
implying that it is indeed transverse. Finally, it is clear that ${\bf H} = d
{\bf B} = d{\bf B}^T$, which means that ${\bf B}^T$ is indeed the physical part of
the two form potential. 

As in the case of gauge free electrodynamics, one can formulate
the theory of Kalb-Ramond fields purely in terms of a {\it physical}
antisymmetric tensor potential $B_{\mu \nu}$ defined by the action 
\begin{eqnarray}
S_{KR}~=~\int d^4x \left( -\frac12 B_{{\cal P} \nu \rho} ~\Box B_{\cal P}^{\nu
    \rho} + J_{\nu \rho} ~B_{\cal P}^{\nu \rho} \right)  ~,
\label{kract}
\end{eqnarray}
where, $\partial^{\mu} B_{{\cal P} \mu \nu} =0 = \partial^{\mu} J_{\mu
\nu}$. 

We once again ask how unique the potential $B_{\cal P \mu \nu}$ is. Observe that both
the field equation and the divergenceless condition remain invariant under a gauge
transformation $B_{\cal P \mu \nu} \rightarrow (B_{\cal P \mu \nu})^{\Lambda}
= B_{\cal P \mu \nu} + 2\partial_{[\mu} \Lambda_{\nu]}$ where $\Lambda_{\mu}$
satisfies the equation $\Box \Lambda_{\mu} - \partial_{\mu} \partial \cdot
\Lambda = 0$. In contrast to the case of the graviton field, it is obvious
that this equation has an infinity of {\it gauge equivalent} solutions, the
equivalence being under $\Lambda_{\mu} \rightarrow \Lambda_{\mu}
+ \partial_{\mu} \omega$ for an arbitrary function $\omega$. Restricting
$\Lambda_{\mu}(\infty) = 0$ is not enough to make it vanish everywhere. We
need to additionally restrict $\partial \cdot \Lambda =0$ everywhere with the
requirement that $\omega(\infty) = const$. This additional restriction appears
necessary in this preliminary investigation to make the two form potential
unique. 

The reason why an identical procedure as for the photon or graviton field
does not suffice to yield a gauge-free formulation of antisymmetric tensor
potentials is because of the aspect of {\it reducibility} of these potentials: the
vectorial gauge parameter of the two form potential itself has a gauge
invariance. Perhaps our approach will need to be somewhat modified to produce
a gauge-free theory of potentials that have a reducible gauge invariance. 

\section{Conclusion}

Generalization of the foregoing approach to Yang Mills theories,
as has already been mentioned, has been achieved in the context of the electroweak
theory where Higgs scalars are assumed to be present \cite{nie, fadd}. In
these papers, a residual $U(1)$ gauge theory corresponding to the Maxwell
theory has been obtained. But since in the foregoing sections a gauge-free
version of the Maxwell theory has been proposed and elaborated upon, one can
assume that this problem is under control, at least at the formulational
stage. A comprehensive study of all quantum properties of such a formulation is
under way and will be reported elsewhere.

For pure Yang Mills theories, the construction of a gauge-free alternative has
not yet been attempted, even though lattice gauge theories represent an
explicitly gauge invariant formulation. A local, gauge-free formulation of
Yang Mills theories is not obviously in contradiction with extant ideas about
colour confinement of quarks and gluons. This gives us the opportunity to
attempt a construction of a physical {\it non-Abelian}
one form in terms of the usual Yang-Mills gauge one form ${\bf A}$ (which takes
values in the Lie algebra of the gauge group ${\cal G}$). 

Defining the holonomy along
the curve $C$ from $y$ to $x$ as $h_{C[y,x]}[{\bf A}] \equiv {\bf P} \exp \int_{C(y,x)} {\bf
A}$, with ${\bf P}$ denoting path ordering, we note that under local gauge
transformations of the gauge potential $[{\bf A}(x)]^{\Omega(x)} =
\Omega(x)^{-1}[{\bf A}(x) + d] \Omega(x)$, where $\Omega \in {\cal G}$
the holonomy variables transform as 
\begin{eqnarray}
 h_{C(y,x)}[{\bf A}^{\Omega}]  = \Omega^{-1}(y) ~h_{C(y,x)}[{\bf A}]~ \Omega(x)
 ~. \label{holo}
\end{eqnarray}
If we choose the point $y \rightarrow \infty$ and require $\Omega(\infty) =
{\cal I}$, eqn. (\ref{holo}) now takes the form 
\begin{eqnarray}
h_{C(\infty, x)}[{\bf A}^{\Omega}] = h_{C(\infty, x)}[{\bf A}]~\Omega(x)
~.\label{grv}
\end{eqnarray}

We now formally define a {\it local} one form potential ${\cal A}(x)$ as
\begin{eqnarray}
{\cal A}(x) \equiv \int {\cal D}C~{\bar {\bf A}}_{C(\infty,x)} 
\end{eqnarray}
where, 
\begin{eqnarray}
{\bar {\bf A}}_{C(\infty,x)} \equiv h_{C(\infty,x)}[{\bf A}] \left( {\bf A} +
  d \right)  (h_{C(\infty,x)}[{\bf A}])^{-1}. ~\label{abar}
\end{eqnarray}
The path integral symbol at this point is formal, and is meant to stand for
some sort of averaging over all paths originating at asymptopia and extending
upto the field point $x$. It is then easy to see that, under gauge
transformations of ${\bf A}$ and
using eqn. (\ref{grv}), 
\begin{eqnarray}
{\cal A}^{\Omega}(x) ~=~ {\cal A}(x) ~. \label{gf}
\end{eqnarray}
What we have not been determined yet is what constraint replaces the
divergencefree condition (\ref{trans}) for the Yang Mills one form ${\cal A}$, so that the
physics of these local gauge-free one forms can be explored
more thoroughly without gauge encumbrances. One also envisages application of these ideas to general
relativity formulated as a gauge theory of Lorentz (or Poincar\'e)
connection. We hope to discuss these and consequent issues elsewhere.  
\vglue .5cm

\noindent {\bf Acknowledgment :} We thank R. Basu, A. Chatterjee, A. Ghosh, B.
Sathiapalan and S. SenGupta for useful discussions.

\end{document}